\documentstyle[12pt]{article}

\title{Transformation Properties of Classical and Quantum Laws
under Some Nonholonomic Spacetime Transformations}
\author{\large A. Pelster
and H. Kleinert\thanks{
Lecture presented by A. Pelster. Emails: pelster@physik.fu-berlin.de;
kleinert@physik.fu-berlin.de;
 URL: http://www.physik.fu-berlin.de/\~{}kleinert. Phone/Fax: 0049/30/8383034
} \\[3mm]
\em Institut f\"ur Theoretische Physik, Freie Universit\"at Berlin \\
\em Arnimallee 14, D-14195 Berlin, Germany}
\date{}
\begin{document}
\maketitle
\begin{abstract}
Nonrelativistic Newton and Schr\"odinger equations
remain correct not only under holonomic but also under nonholonomic
transformations of the spacetime coordinates.
Here we study the properties
of transformations which
are holonomic in the space coordinates while additionally
tranforming the
time in a path-dependent way.
This makes them nonholonomic in spacetime.
The resulting transformation formulas of physical quantities
establish relations between different physical systems.
Furthermore we point out certain
differential-geometric features of these relations.
\end{abstract}
\section{Introduction}

Based on the pioneering work of Poincar\'e$^{1}$ and Sundman$^{2}$
in celestial mechanics,
Kustaanheimo and Stiefel$^{3}$ solved the Kepler problem
in 1965 by using a local transformation of
space and time which possesses no global counterpart.
This nonintegrable transformation regularizes
the singular three-dimensional Kepler problem by mapping it to the
four-dimensional harmonic oscillator. Motivated by
the discovery of the dynamical symmetry group O(4,2)$^{4}$,
Duru and Kleinert$^{5,6}$ showed in 1979
that these transformations can also be implemented
in path integrals. Generalizations of this
procedure
have led to relations beween Green functions
of many different
quantum systems, thereby producing previously unknown solutions$^{7}$.\\

Here we discuss differential-geometric
properties of a special class of these
transformations, in which the spatial part of the transformations
is holonomic.
This excludes the treatment of the Kepler problem in three dimensions,
but
admits all  one-dimensional DK transformations (see Chaper 14 of Ref.~7).

Consider
the movement of a point mass $m$
on a $D$-dimensional Riemann manifold under the influence of external
potentials. Both classical
and quantum mechanical equations are form invariant
under ordinary {\em holonomic\/}
space transformations. Locally, some
 initial space coordinates
$q^{\,i}$
are changed into
$Q^{\,\lambda}$~($i,\lambda = 1 ,
\ldots, D$) according to
\begin{eqnarray}
\label{STL}
d  q^{\,i} \, = \, e^{\,i}_{\,\,\,\lambda} ( Q ) \, d Q^{\,\lambda} \, ,
\end{eqnarray}
where the coefficients $e^{\,i}_{\,\,\,\lambda} ( Q	)$ obey the
integrability conditions of Schwarz
\begin{eqnarray}
\label{I}
\partial_{\mu} \, e^{\,i}_{\,\,\,\lambda} ( Q ) \, - \,
\partial_{\lambda} \, e^{\,i}_{\mu} ( Q ) \, = \, 0 \, .
\label{schw}\end{eqnarray}
This is why (\ref{STL}) is integrable and possesses the global form
\begin{eqnarray}
\label{ST}
q^{\,i} \, = \, q^{\,i} ( Q ) \, .
\end{eqnarray}
In the following we investigate
what happens to this form invariance
if the time $t$ is transformed under a local transformation
to a new time $s$ via
\begin{eqnarray}
\label{TT}
\frac{d t}{d s} \, = \, f ( Q ) \, ,
\end{eqnarray}
where the positive but otherwise arbitrary function $f ( Q )$ depends
on the final space coordinates.
The combined local space and time transformation
\begin{eqnarray}
\left( \begin{array}{@{}c} d t \\ \, d q^{\,i} \end{array} \right) \, = \,
\left( \begin{array}{@{}cc} f ( Q ) & 0 \\ 0 & e^{\,i}_{\,\,\,\lambda}
( Q ) \end{array} \right) \, \left( \begin{array}{@{}c} d s \\
\,d Q^{\,\lambda}
\end{array} \right)
\label{spt}\end{eqnarray}
is characterized by
 transformation coefficients $E^I{}_ \Lambda$ ($I, \Lambda=0,\dots , D$)
which no longer
fulfill the time part of the spacetime
extension of the integrability condition (\ref{schw}), making
 (\ref{TT}) nonholonomic in spacetime.\\

\section{Classical Mechanics}

Consider a trajectory $q^{\,i} ( t )$
of a point mass $m$ in a $D$-dimensional Riemann
manifold with metric $g^{({\rm i})}_{ij} ( q )$ under the influence
of a vector potential $A^{({\rm i})}_{\,i} ( q )$ and a scalar
potential $V^{({\rm i})} ( q )$. This
problem may be solved
by integrating the equations of motion, but
it may often be simplified by
subjecting the
system to nonholonomic spacetime transformations
of the type (\ref{ST}) and (\ref{TT}).
They relate
the trajectory $q^{\,i} ( t )$ to some
trajectory $Q^{\,\lambda} ( s )$ in a
space with metric $g^{({\rm f})}_{\lambda\mu}
( Q )$, vector potential $A^{({\rm f})}_{\,\lambda} ( Q )$
and scalar potential $V^{({\rm f})} ( Q )$ as is depicted in Fig. 1.\\

The holonomic space transformation (\ref{ST}) changes the metric
in accordance with its tensor character.
The
nonholonomic time transformation
(\ref{TT}) turns out to produce an additional conformal factor$^{8}$.
Together we end up with
\begin{eqnarray}
\label{MT3}
g^{({\rm f})}_{\lambda\mu} ( Q )&  = & f ( Q ) \,
e^{\,i}_{\,\,\,\lambda} ( Q ) \, e^{\,j}_{\,\,\,\mu} ( Q )\,
g^{({\rm i})}_{ij} \left( q ( Q ) \right) \, .
\end{eqnarray}
The vector potential is transformed by the holonomic
space transformation (\ref{ST}) as usual.
In contrast to the metric, it
is unchanged by the nonholonomic time transformation (\ref{TT}) as it is
coupled to the velocity. Together we obtain
\begin{eqnarray}
\label{CVP}
A^{({\rm f})}_{\,\lambda} ( Q ) \, = \, e^{\,i}_{\,\,\,\lambda} ( Q ) \,
A^{({\rm i})}_{\,i} ( q ( Q ) ) \, .
\end{eqnarray}
If the energy of the orbit $q^{\, i} (t)$ is denoted by $E^{({\rm i})}$,
the
transformation of the scalar potential is$^8$
\begin{eqnarray}
\label{SPT4}
V^{({\rm f})} ( Q ) \, = \, f ( Q ) \, \left\{ V^{({\rm i})} \left( q ( Q )
\right) \, - \, E^{({\rm i})} \right\} \, .
\label{trsp}\end{eqnarray}
Note that the initial energy $E^{({\it i})}$
plays the role of a strength parameter in an additional potential proportional
to
$-f(Q)$.
The transformation of the trajectory is
given by
\begin{eqnarray}
\label{TTRU}
q^{\,i} ( t ) \, = \, q^{\,i} \left( Q ( s ) \right)\, ,
\end{eqnarray}
where  the relation  between both time coordinates $t$ and $s$
is obtained by integrating
(\ref{TT}) along the trajectory:
\begin{eqnarray}
\label{ZT}
t \, = \, \int \limits^{s}_{0}
d s \, f \left( Q ( s ) \right).
\end{eqnarray}
\begin{figure}[t]
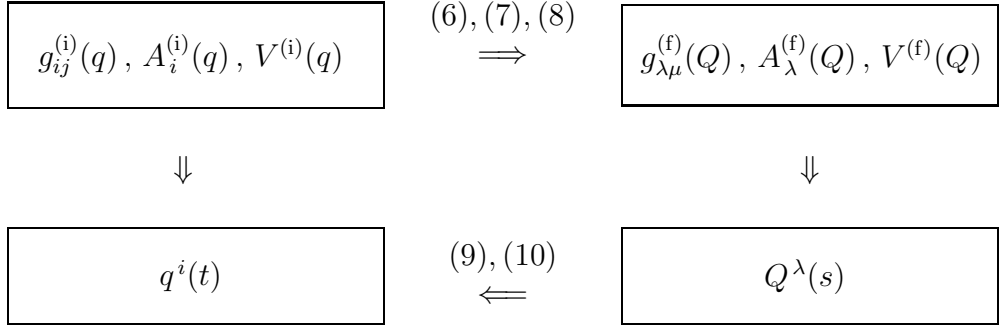

\vspace*{-0.2cm}
\begin{center}
\begin{tabular}{@{}ccc}
\fbox{
\begin{minipage}{4.5cm}
\begin{center}
\vspace{0.3cm}
$g^{({\rm i})}_{ij} ( q ) \, , \, A^{({\rm i})}_{\,i} ( q ) \, , \,
V^{({\rm i})} ( q )$
\vspace{0.3cm}
\end{center}
\end{minipage} }
& $\begin{array}{c}
(\ref{MT3}), (\ref{CVP}), (\ref{SPT4}) \\ \Longrightarrow \\ \mbox{}
\end{array}$
& \fbox{
\begin{minipage}{4.5cm}
\begin{center}
\vspace{0.2cm}
$g^{({\rm f})}_{\lambda\mu} ( Q ) \, , \, A^{({\rm f})}_{\,\lambda} (  Q )
\, , \, V^{({\rm f})} ( Q )$
\vspace{-0.1cm}
\end{center}
\end{minipage}  }
\\
\hspace{0.2cm} & \hspace{0.2cm} &  \hspace{0.2cm} \\
$\Downarrow$ \,\,\,  & \hspace{1cm}  &  $\Downarrow$ \\
\hspace{0.2cm} & \hspace{0.2cm}  & \hspace{0.2cm} \\
\fbox{
\begin{minipage}{4.5cm}
\begin{center}
\vspace{0.3cm}
$q^{\,i} ( t )$
\vspace{0.3cm}
\end{center}
\end{minipage} }
& $\begin{array}{c}
(\ref{TTRU}), (\ref{ZT}) \\ \Longleftarrow \end{array}$
& \fbox{
\begin{minipage}{4.5cm}
\begin{center}
\vspace{0.3cm}
$Q^{\,\lambda} ( s )$
\vspace{0.3cm}
\end{center}
\end{minipage} } \\
\end{tabular}
\end{center}
 \caption{\it Nonholonomic space and time transformation relating
the metrics,
external potentials and trajectories
of two
classical systems with each other.}
\vspace*{0.2cm}
\end{figure}
\begin{figure}[tbh]
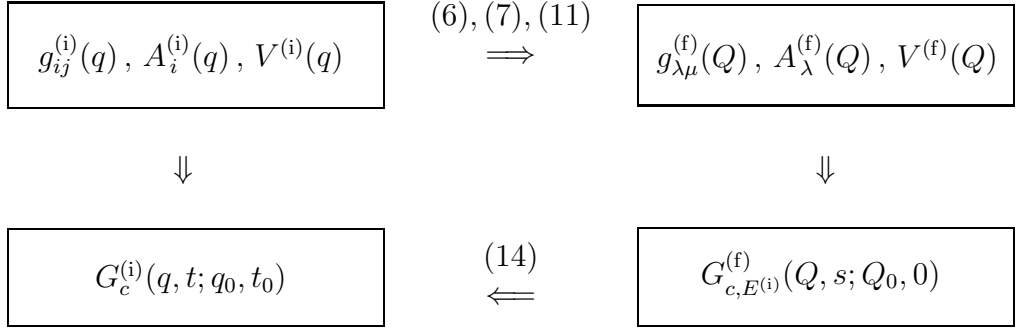

\vspace*{-0.2cm}
\begin{center}
\begin{tabular}{@{}ccc}
\fbox{
\begin{minipage}{4.5cm}
\begin{center}
\vspace{0.3cm}
$g^{({\rm i})}_{ij} ( q ) \, , \, A^{({\rm i})}_{\,i} ( q ) \, , \,
V^{({\rm i})} ( q )$
\vspace{0.3cm}
\end{center}
\end{minipage} }
& $\begin{array}{c}
(\ref{MT3}), (\ref{CVP}), (\ref{POTI20}) \\ \Longrightarrow \\ \mbox{}
\end{array}$
& \fbox{
\begin{minipage}{4.5cm}
\begin{center}
\vspace{0.2cm}
$g^{({\rm f})}_{\lambda\mu} ( Q ) \, , \, A^{({\rm f})}_{\,\lambda} ( Q )
\, , \,
V^{({\rm f})} ( Q )$
\vspace{-0.1cm}
\end{center}
\end{minipage}  }
\\
\hspace{0.2cm} & \hspace{0.2cm} &  \hspace{0.2cm} \\
$\Downarrow$ \,\,\,  & \hspace{1cm}  &  $\Downarrow$ \\
\hspace{0.2cm} & \hspace{0.2cm}  & \hspace{0.2cm} \\
\fbox{
\begin{minipage}{4.5cm}
\begin{center}
\vspace{0.3cm}
$G^{({\rm i})}_c ( q , t ; q_0 , t_0 )$
\vspace{0.3cm}
\end{center}
\end{minipage} }
& $\begin{array}{c}
(\ref{PROTRA7}) \\ \Longleftarrow \end{array}$
& \fbox{
\begin{minipage}{4.5cm}
\begin{center}
\vspace{0.2cm}
$G^{({\rm f})}_{c,E^{({\rm i})}} ( Q , s ; Q_0 , 0 )$
\vspace{0.25cm}
\end{center}
\end{minipage} } \\
\end{tabular}
\end{center}
 \caption{{\it Nonholonomic space and time transformation relating
the metrics,
external potentials and time evolution amplitudes
of two
quantum systems with each other.}}
\vspace*{0.2cm}
\end{figure}

\section{Quantum Mechanics}

Let the quantum mechanical system of the
point mass $m$ in a
space  with
metric $g^{({\rm i})}_{ij} ( q )$, vector potential
$A^{({\rm i})}_{\,i} ( q )$, and scalar potential $V^{({\rm i})} ( q )$
be described by the time evolution amplitude
$G^{({\rm i})}_c ( q , t ; q_0 , t_0 )$.
This can be calculated by solving a Schr\"odinger equation
or evaluating a path integral. The nonholonomic spacetime
transformations (\ref{ST}) and (\ref{TT})
 make it possible to obtain
$G^{({\rm i})}_c ( q , t ; q_0 , t_0 )$ from an analogous amplitude
in the transformed system. The respective tranformation properties
of the physical quantities are illustrated in Fig. 2.\\

The quantum mechanical transformations of the metric and the
vector potential turn out to coincide with the classical ones (\ref{MT3})
and (\ref{CVP}).
However, in contrast to the classical problem,
the scalar potential
of the transformed system contains an additional
quantum correction term proportional to $\hbar ^2$
\begin{eqnarray}
& &\!\!\!\!\!\!\!\!\!\!\!\!\!\!\!\!\!\!
V^{({\rm f})} ( Q ) \, = \, f ( Q ) \, \left\{ V^{({\rm i})} ( q ( Q ) ) \, -
\,
E^{({\rm i})} \right\} \, + \, \frac{\hbar^2}{m} \left\{ \frac{2 - D}{8} \,
\Gamma^{({\rm f})\lambda\mu}_{\lambda} ( Q ) \,
\frac{\partial_{\mu} f ( Q )}{f ( Q )} \right. \nonumber \\
\label{POTI20}
& & ~~+~ g^{({\rm f})\lambda\mu} ( Q ) \left[ \frac{(D - 2)(D - 6)}{32} \,
\frac{\partial_{\lambda} f ( Q ) \partial_{\mu} f ( Q )}{f ( Q )^2}
\left. + \, \frac{D - 2}{8}\,
\frac{\partial_{\lambda}\partial_{\mu} f ( Q )}{f ( Q )}
\right] \right\} ,
\end{eqnarray}
where
$\Gamma^{({\rm f})\,\,\,\nu}_{\lambda\mu} ( Q )$
is the
Christoffel symbol
associated with the metric
$g^{({\rm f})}_{\lambda\mu} ( Q )$.
Note that the quantum corrections
happen to be absent for $D = 2$ and reduce
 for $D=1$ to known
results$^{7}$.\\

An analysis$^8$  of the differential geometric properties
of the nonholonomic spacetime transformations
(\ref{ST}) and (\ref{TT})
 in the framework
of Riemann-Cartan differential geometry$^{9,10}$
reveals
that
the quantum corrections
can be expressed in terms of curvature and torsion of the
transformed spacetime.
If $R^{({\rm f})} ( Q )$, $\overline{R}^{({\rm f})} ( Q )$
and $S^{({\rm f})}_{\lambda} ( Q ) \equiv S^{({\rm f})}_{ \lambda\mu}{}^\mu
( Q )$
 denote Cartan's curvature scalar, Riemann's
curvature scalar, and contracted Cartan's torsion tensor, respectively, one
has instead of
(\ref{trsp})
\begin{eqnarray}
 V^{({\rm f})} ( Q )
=  f ( Q ) \, \big\{ V^{({\rm i})} \left( q ( Q ) \right)
\, - \, E^{({\rm i})}\big\}+
 V^{({\rm qu})} ( Q )
\end{eqnarray}
with the quantum correction
\begin{eqnarray}
& &  \!\!\!\!\!\!\!\!\!\!\!\!\!\!\!\!\!\!\!\!
 V^{({\rm qu})} ( Q )
=  f ( Q ) \Bigg\{
 \frac{\hbar^2}{m} \, \left[ \frac{4 - D^2}{8} \,
\left( S^{({\rm f})} ( Q ) \right)^2
 \, + \, \frac{D - 2}{16}
\, \left( R^{({\rm f})} ( Q ) \, - \, \overline{R}^{({\rm f})}
( Q ) \right) \right] \Bigg\} \, .
\label{WEGE}
\end{eqnarray}
It is not astonishing that
the paths in quantum mechanics
whose quantum fluctuations probe the neighborhood of the classical orbits
are sensitive
to curvature and torsion
of
spacetime.
\\

The relation between the
initial time evolution amplitude
$G^{({\rm i})}_c ( q , t ; q_0 , t_0 )$
and the corresponding amplitude
$G^{({\rm f})}_{c, E^{({\rm i})}} \left( Q ( q ) , s ; Q ( q_0 )  , 0 \right)$
in the transformed
system
is given by the DK transformation$^8$ which essentially contains two
Fourier transformations:
\begin{eqnarray}
\!\!\!\!\!\!\!\!G^{({\rm i})}_c ( q , t ; q_0 , t_0 ) \, &= &\, \left[ f \left(
Q ( q ) \right) \,
f \left( Q ( q_0 ) \right)
\right]^{\frac{2 - D}{4}} \,
\int\limits_{- \infty}^{+ \infty} \, \frac{d E^{({\rm i})}}{2 \pi \hbar} \,
\int\limits_{0}^{+ \infty} \, d s
\nonumber\\
\label{PROTRA7}
&&\times \, \exp \left\{ - \, \frac{i}{\hbar} \, E^{({\rm i})} \, ( t - t_0 )
\right\}
\, G^{({\rm f})}_{c, E^{({\rm i})}} \left( Q ( q ) , s ; Q ( q_0 )  , 0 \right)
\, .
\end{eqnarray}

\section{Outlook}

It will be interesting to extend this dicussion to
allow for nonholonomic space
transformations
which inlude the Kustaanheimo-Stiefel case (which by themselves are discussed
in
Kleinert's lecture and in Ref. 11),
and to
completely general nonholonomic spacetime transformations.

\section{References}

\begin{tabular}{lp{14.05cm}}
{}~1. &  H. Poincar\'{e}: {\it Sur le Probl\`{e}me des
Trois Corps et les \'{E}quations de la Dynamique;}
Acta math. {\bf 13}, 1--271 (1890)\\
{}~2. &  K.F. Sundman: {\it M\'{e}moire sur le Probl\`{e}me
des Trois Corps;} Acta math. {\bf 36}, 105--179 (1913)\\
{}~3. & P. Kustaanheimo, E. Stiefel: {\it Perturbation Theory of
Kepler Motion based on Spinor Regularization;} J. Reine Angew.
Math. {\bf 218}, 204--219 (1965)\\
{}~4. & H. Kleinert: {\it Group Dynamics of Elementary
Particles;} University of Colorado Thesis, Fortschritte der Physik
{\bf 16}, 1--74 (1968)\\
{}~5. & I.H. Duru, H. Kleinert: {\it Solution of the Path
Integral for the H-Atom;} Phys. Lett. {\bf B 84}, 185--188 (1979)\\
{}~6. & I.H. Duru, H. Kleinert: {\it Quantum Mechanics of
H-Atom from Path Integrals;} Fortschr. Phys. {\bf 30},
401--435 (1982)\\
{}~7. & H. Kleinert: {\it Path Integrals in Quantum Mechanics,
Statistics and Polymer Physics;} World Scientific, Second Edition (1995)\\
{}~8. & A. Pelster: {\it Zur Theorie und Anwendung nichtintegrabler
Raum-Zeit-Trans\-for\-ma\-tionen in der klassischen Mechanik und in der
Quantenmechanik;} Dissertation, Universit\"at Stuttgart, Shaker-Verlag (1996)\\
{}~9. &
J.A. Schouten: {\it Ricci Calculus};
Springer-Verlag, Berlin  (1954)\\
$\!$10. &H. Kleinert:
     {\it Gauge Fields in Condensed Matter\/},
     Vol.\ II \,  Part IV ({\it Gravity with Torsion\/});
     World Scientific Publishing Co., Singapore (1989)\\
$\!$11. & H. Kleinert: {\it Classical and Fluctuating Paths in Spaces
with Curvature and Torsion;} Lecture presented at the {\it 5th International
Conference on Path Integrals from mev to Mev, Dubna, May 27--31, 1996}
\end{tabular}

\end{document}